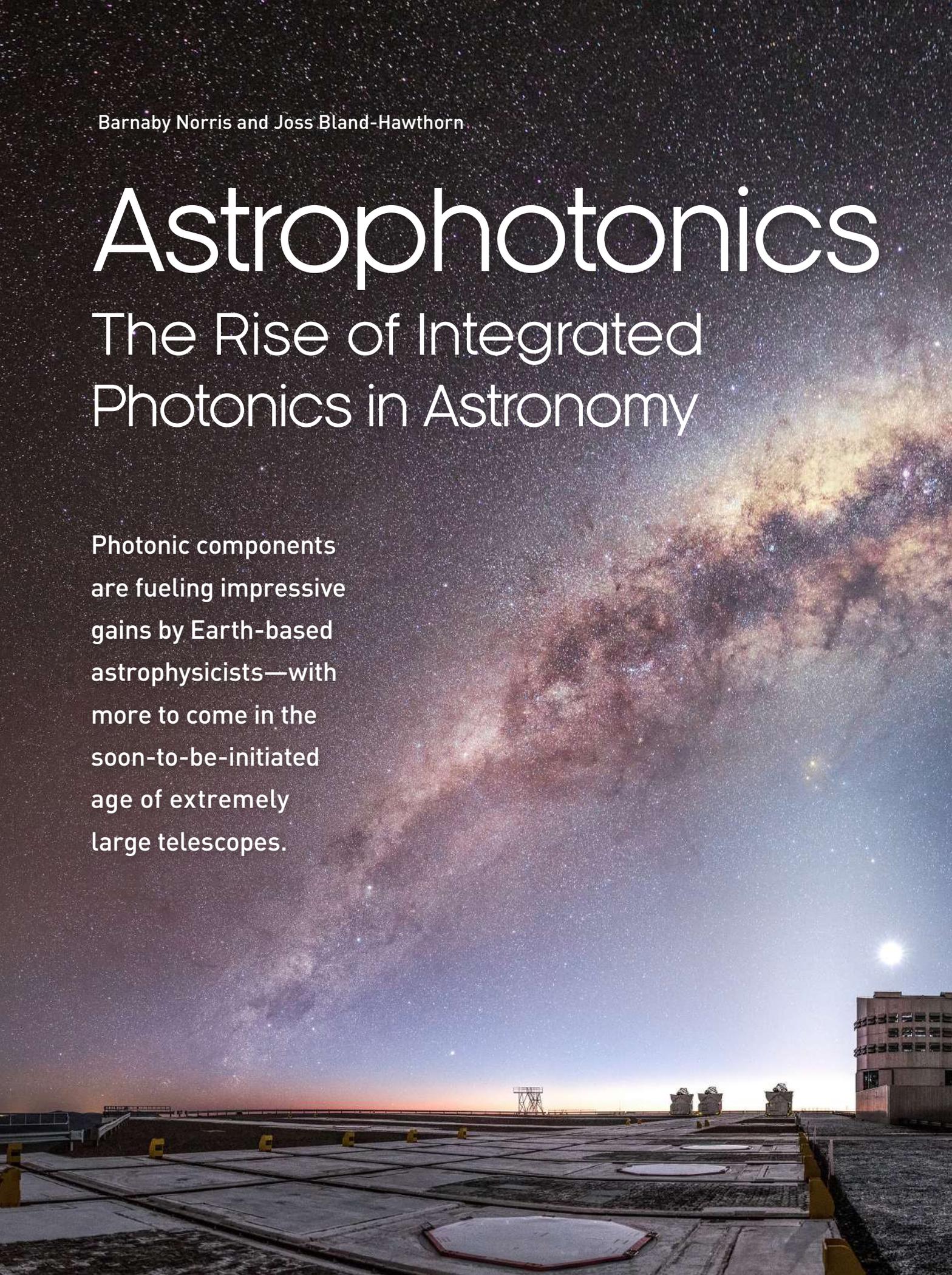

Barnaby Norris and Joss Bland-Hawthorn

# Astrophotonics
## The Rise of Integrated Photonics in Astronomy

Photonic components are fueling impressive gains by Earth-based astrophysicists—with more to come in the soon-to-be-initiated age of extremely large telescopes.

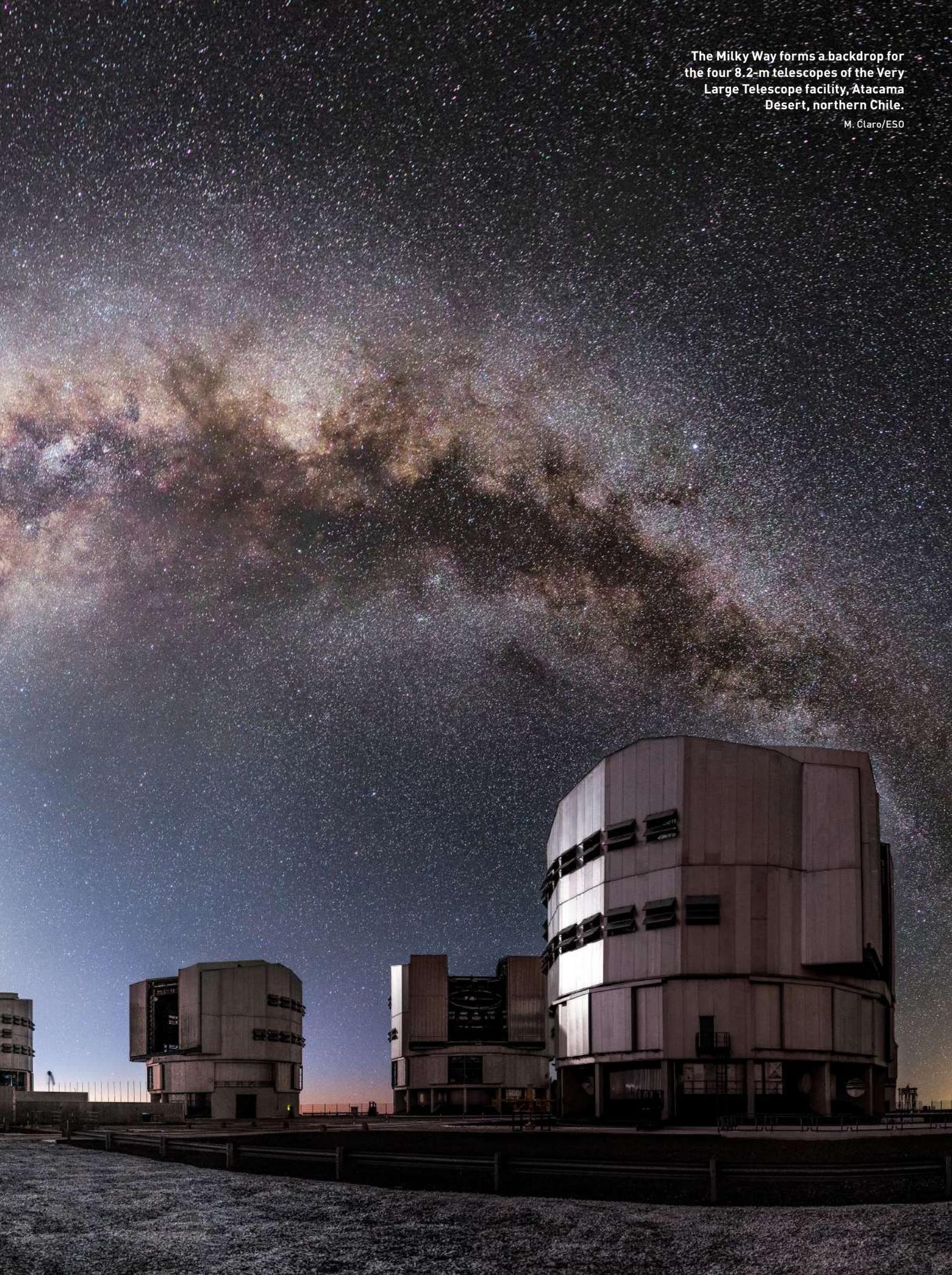

The Milky Way forms a backdrop for the four 8.2-m telescopes of the Very Large Telescope facility, Atacama Desert, northern Chile.
M. Claro/ESO

Astronomy, one of the oldest scientific endeavors, has always sought to take advantage of new technology. After the spectacle maker Hans Lippershey attempted to patent his refracting telescope, Galileo moved quickly to build his own device to study the night sky. The rest, as they say, is history.

Some three and one-half centuries after Galileo's death, astronomy continues to rapidly evolve, now asking questions over vast, and vastly different, scales. In the nearby universe, we search for biosignatures on extrasolar planets as possible evidence for life beyond our solar system. We track the motions of billions of stars throughout the Milky Way and around the black hole at the galaxy's center. At the other extreme, we now observe at distances where the time of light almost equals the age of the universe, glimpsing galaxies as they appeared billions of years ago.

To make these observations, modern astronomers commonly use optical solutions in the form of large, often segmented mirrors to bring light to a focus. The focused light then goes to an astronomical instrument defined by optical components such as lenses, polarizers and gratings, where the processed light is re-imaged onto a detector. Today, we are embarking on an era of extremely large telescopes (ELTs) with mirror diameters from 25 to 40 m, polished and aligned to a precision of around 15 nm across the whole surface. ELTs will allow us to see faint planets around nearby stars and the most distant galaxies in the universe—and could make the period from 2025 to 2050 a golden era for astronomy.

The biggest monolithic telescopes, such as the ELTs, will always be on Earth rather than in space. And that presents a major challenge: the distortion of celestial light by our turbulent atmosphere. Telescopes are becoming larger to gather more light, and more sophisticated to correct for atmospheric interference. Recently, astronomers have come to recognize the benefits of integrated photonics, often in combination with optical systems, in achieving that quest. Here, we explore some of the recent and future advances in Earth-based astronomy made possible by integrated photonics.

## Laser guide stars

In recent decades, astronomers have learned to generate "artificial stars" at an altitude of 90 km using powerful lasers. These coherent sources, tuned to the 589-nm sodium line, excite high-altitude sodium atoms in Earth's mesosphere, which serve as known reference points; any wavefront distortions in the reflected light must thus be due to the intervening atmosphere. Adaptive optics (AO) allow astronomers to monitor the atmosphere's rapidly changing distortions, and to deform internal mirrors at a rate of several kHz to compensate for the wavefront distortion.

Bright, natural stars can also be used as guide stars, but this relies on the existence of a suitable bright star directly within the field of astronomical interest. (Stars are intrinsically incoherent light sources but at extremely large distances most appear spatially coherent—this is the van Cittert-Zernike theorem in action.) Laser guide stars, on the other hand, allow even regions containing only faint sources to be observed.

In recent years, fiber lasers have replaced bulk lasers for generating these artificial stars. Typical systems, consisting of a frequency-doubled diode laser and Raman fiber amplifier, provide output powers of around 20 W. These systems avoid the problems of maintaining alignment in bulk lasers in the hostile, remote mountaintop environments where big telescopes reside.

### Integrated optical devices for astrophotonics

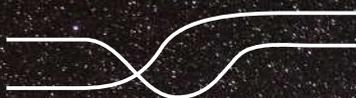

**Single-mode waveguides** transport light between components with precisely matched optical path difference, maintaining coherence and performing spatial filtering.

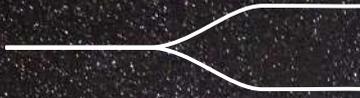

**Y-splitters/junctions** can split individual telescopes/sub-pupils for later recombination, or combine beams with a single interference output.

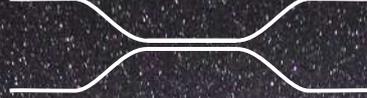

**Directional couplers** combine beams with two interference outputs (dark and light), maintaining all flux.

Centuries after Galileo's death, astronomy continues to evolve, asking questions over vast, and vastly different, scales.

AO systems are now fully integrated into the design of telescopes, and can compensate for the bulk of the distortion induced by the turbulent atmosphere. But there are always residual, uncompensated effects that must be handled downstream within the optical instrument located at the telescope focus. The versatility of photonics gives us many ways to manipulate light with a view to forming an undistorted image.

### Coupling light

Photonics allows for some remarkably simple but effective operations that have no other counterpart in optics. For example, if we place a single-mode fiber (or a waveguide within an integrated-photonic chip) at the telescope focus, we can couple some light into it, albeit inefficiently. For light to couple into the fiber, it must have a wavefront whose wave vector is perpendicular to the fiber's front face. By definition, the fiber "cleans" the input beam, allowing only coherent light to couple into the fiber. Since the fiber supports only the fundamental mode, higher-order spatial information is filtered out (at the cost of injection efficiency). The fiber thus acts as the perfect spatial filter, an invaluable property for interferometry and spectroscopy.

Ideally, the beam would have a Gaussian cross-section with a diameter matched to that of the fiber core. AO systems come close to producing a diffraction-limited circular beam for optimal injection, but the results are never perfect. If the corrected beam is smeared spatially, the coupling will be poor, leading to low overall throughput in the astronomical instrument. Astrophotonics provides

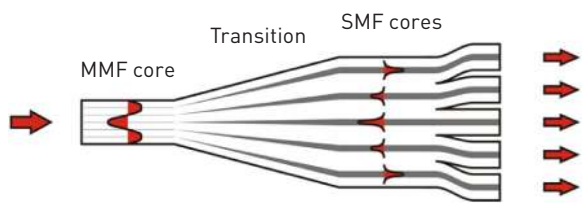

### The photonic lantern

A multimode fiber device consisting of an array of identical single-mode fiber cores, the photonic lantern allows the light signal captured by a telescope's multimode fiber to be channeled into an array of output single-mode waveguides.
Adapted from S.G. Leon-Saval et al., Opt. Lett. **30**, 2545 (2005)

another solution to this quandary: the photonic lantern, a multimode fiber device consisting of an array of identical single-mode fiber cores.

The lantern allows telescopes to use a large-diameter multimode fiber at the telescope focus that receives essentially all of the focused light. If $N$ spatial (unpolarized) modes couple into the photonic lantern, these can be coupled efficiently into $N$ output single-mode waveguides—a device referred to as a lantern waveguide. Generally, there is no one-to-one mapping between a specific input mode and output waveguide; the information is shared across all outputs. Thus beam-smeared incoherent light can be converted to single-mode propagation along a set of waveguides that can be used as inputs to other photonic actions—for example, array waveguide gratings (AWGs) or Bragg gratings. All of this functionality can now be integrated into a 3-D photonic device.

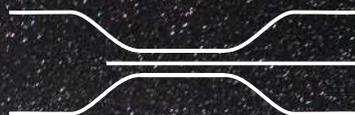

**Tri-couplers** and other **multi-input, multi-output couplers** allow more complex beam combinations to be performed.

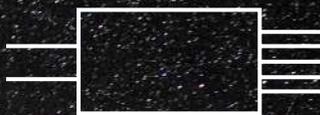

**Multimode interference (MMI) couplers** use a multimode propagation region, rather than evanescent coupling, to perform complex input-output arrangements.

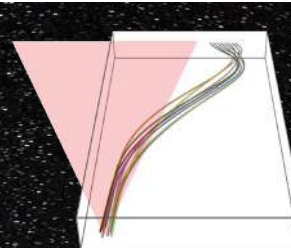

**Pupil-remappers** have a 3-D arrangement of waveguides to convert sub-pupils from a single telescope pupil into a linear array for combination, while maintaining optical path length.

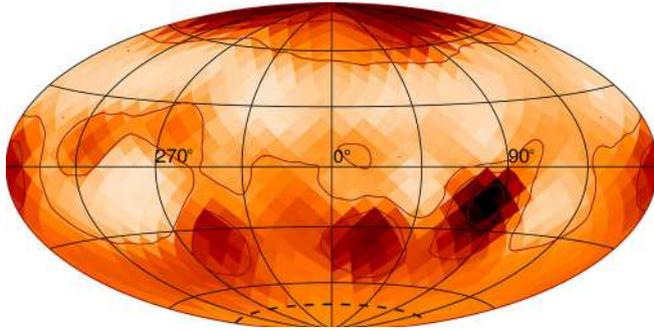

In 2016, international astronomers using the MIRC interferometer at the Mt. Wilson Observatory made the first detailed observation of "starspots" on the surface of a star outside of our solar system.

Rachael Roettenbacher and John Monnier

### Interferometry

Instead of, or in addition to, correcting the input beam up front with adaptive optics, another approach is to receive the badly aberrated signal from the telescope, and then disentangle the true astrophysical signal from the corrupted data via complex waveguide architectures and advanced data-processing techniques. These interferometric methods directly measure the spatial coherence properties of the starlight. From this, several observable quantities can be extracted that are a function only of the image of the distant astronomical source, and not of the atmospheric seeing. For example, since the turbulent atmosphere generally affects the phase but not the amplitude of the incoming wavefront, the Fourier power spectrum of the signal divulges the spatial frequencies present in the astronomical image without corruption by the atmosphere.

To accomplish this, separate regions of the telescope pupil are injected into multiple waveguides inscribed within a photonic chip. Next, light from each of these sub-pupils can be split multiple times using photonic splitter elements in the chip. Then the light from each sub-pupil is interfered with every other via directional-coupler elements within the chip (wherein two waveguides are brought into close proximity allowing their evanescent fields to overlap) or via multimode interference couplers. Cascading a suitable series of splitters and couplers makes it possible to produce a set of outputs that encodes, in their intensities, a description of the spatial coherence structure of the incoming light; algorithms can use that intensity encoding to reconstruct an image or fit astrophysical models.

Critically, the optical path lengths between all waveguides and photonic elements must be matched at the subwavelength level, to maintain temporal coherence. Integrated-photonic chips are ideally suited to this purpose; their monolithic design allows extremely stable and precise arrangement of waveguides and components, to precisions of a few hundred nanometers, and can be produced in 2-D designs using photolithography or with freeform 3-D designs using the femtosecond laser direct-write process, enabling even greater complexity.

### Putting telescopes together

Another powerful but complex approach is to combine the signal from multiple telescopes. By leveraging photonic devices combined with advanced free-space optics, groundbreaking instruments such as the GRAVITY instrument at the Very Large Telescope Interferometer (VLTI), run by European Southern Observatory (ESO), can be realized. The instrument's awesome scale and complexity emerge in an ESO virtual fly-through movie (www.eso.org/public/videos/eso1622b).

The input beams from the four VLTI telescopes are AO-corrected, and the signals from all foci are phase-referenced with respect to each other and a bright source in the field. In this case, the longest baseline (the distance between sub-pupils, or telescopes) can be hundreds of meters, which dictates the diffraction-limited resolution of the instrument once the observing wavelength is specified. Last year, the GRAVITY team published spectacular results of a fast-moving star, S0-2, zipping past the supermassive black hole at the galactic center, with an orbit fully in accord with Einstein's general theory of relativity.

Impressively, the Michigan Infrared Beam Combiner (MIRC), an interferometer tied to the CHARA facility at Mount Wilson, Calif., USA, brings together the light of six separate telescopes, with a longest baseline of 330 m and a spatial resolution of about 0.5 milli-arcseconds at optical wavelengths. In 2016, that resolution allowed the first direct imaging of "starspots" on the surface of a distant star, ζ Andromedae, 180 light years away.

The key to the success of ultra-high-resolution instruments such as GRAVITY and MIRC lies in the combination of advanced bulk-optical engineering and photonic technologies. Light from separate telescopes must traverse beamlines hundreds of meters long to a central laboratory, where optical delay lines dynamically compensate for the optical path distance arising from the fact one telescope is infinitesimally closer to the star than another. This is accomplished by variable optical delay lines of hundreds of meter lengths,

By leveraging photonic devices and advanced free-space optics, groundbreaking instruments such as GRAVITY at the Very Large Telescope Interferometer can be realized.

measured to sub-micron precision. Multiple advanced AO systems clean up the wavefront from each telescope in preparation for injection into the single-mode photonics. Finally, the light enters the photonic chip that lies at the heart of the instrument, wherein the spatial coherence properties are analysed. The chip outputs must then be imaged at high speeds—to "freeze" the variable atmosphere—at wavelengths ranging from visible to up to 5 µm.

## Nulling interferometry

In the interferometric methods discussed so far, the light and associated photon noise from all objects in the field of view are present at all outputs. This poses a challenge for high-contrast imaging, such as imaging of an exoplanet, which may be $10^4$–$10^{10}$ times fainter than its parent star. The star's photon noise thus completely obliterates the faint signal from the planet.

To address this, an interferometric instrument can be designed so that the light from the parent star is destructively interfered with itself, or "nulled," and only the uncontaminated light from the planet emerges from the outputs of interest. Making this happen requires precise control of optical delays, splitting ratios and other parameters—ideally suited to the precision of a photonic chip. This novel technology is being deployed on-sky, as in the GLINT instrument at the Subaru telescope.

Throughout all of these interferometric applications, the intrinsic spatial-filtering effect of single-mode waveguides plays a key role. The process of making accurate measurements or reconstructing an image from interferometric data relies on having a clean, high-contrast interference pattern, such as that formed by a set of arbitrarily small slits. In other words, while robust against atmosphere-induced wavefront error occurring between telescopes or sub-apertures, it is not so robust against wavefront error occurring within a given sub-aperture.

The use of single-mode waveguides solves this, removing higher order spatial structure within each sub-element and yielding vastly improved precision. In the case of nulling interferometry, the ideal wavefront

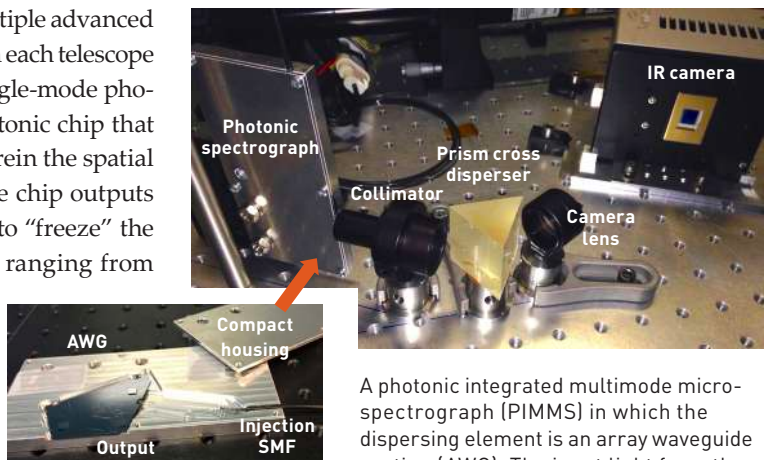

A photonic integrated multimode microspectrograph (PIMMS) in which the dispersing element is an array waveguide grating (AWG). The input light from the Subaru telescope focus (Mauna Kea Observatory, Hawaii, USA) is cleaned by the adaptive-optics system (SCExAO) before being fed to the instrument.
Courtesy of Nick Cvetojevic, Observatoire de la Côte d'Azur, France

resulting from the single-mode waveguides allows the darkest possible null (destructive interference) to be produced, maximally removing the contaminating starlight and revealing even the faintest planetary companions.

Using combinations of these techniques, interferometers have recently tracked stars close to the black hole Sgr A* that lies at the center of the Milky Way, probed the dusty chaos surrounding distant stars and caught the dance of planet formation in-the-act in far-away solar systems.

## Microspectrographs

The primary instrument at most astronomical observatories is the spectrograph. And this instrument, too, is an area that is benefiting from a breakthrough integrated-photonic application developed at the University of Sydney: the photonic integrated multimode microspectrograph (PIMMS).

For a given spectroscopic resolution of an instrument that's not diffraction limited, the instrument's size scales with the slit width, which in turn scales with the diameter of the telescope mirror. Modern spectrographs can be up to 6 m long, and will only grow larger with the giant mirrors of the next-generation ELTs. The PIMMS sidesteps the need for these huge optical instruments.

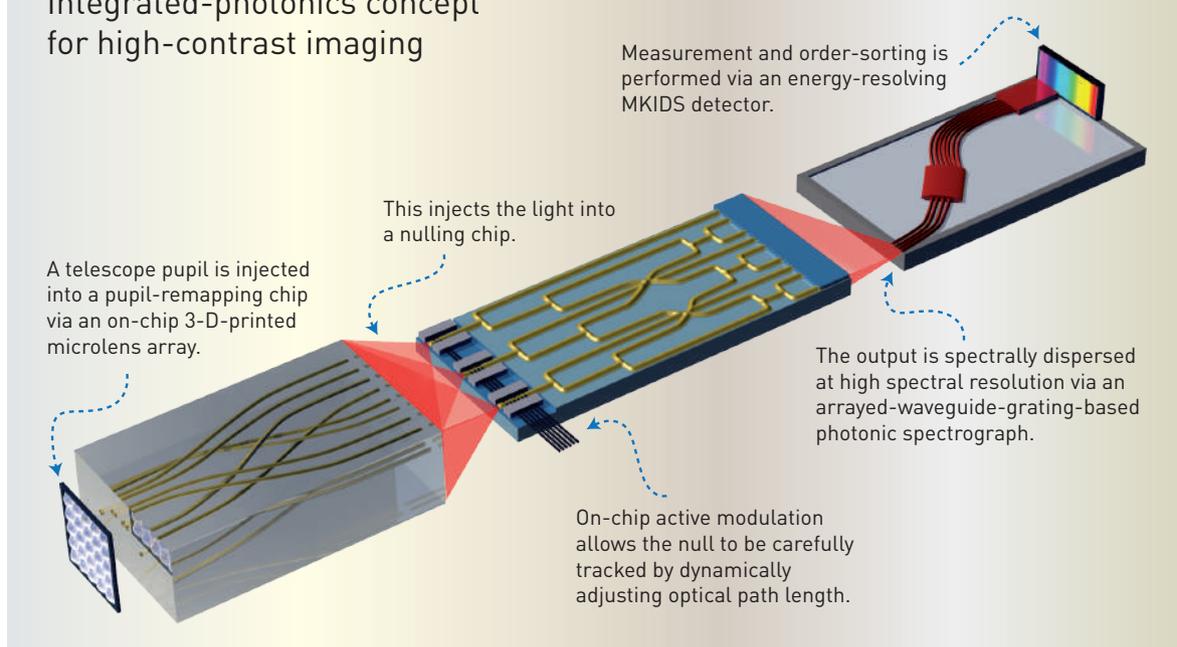

Integrated-photonics concept for high-contrast imaging

A telescope pupil is injected into a pupil-remapping chip via an on-chip 3-D-printed microlens array.

This injects the light into a nulling chip.

Measurement and order-sorting is performed via an energy-resolving MKIDS detector.

The output is spectrally dispersed at high spectral resolution via an arrayed-waveguide-grating-based photonic spectrograph.

On-chip active modulation allows the null to be carefully tracked by dynamically adjusting optical path length.

Illustration by Phil Saunders

Using the photonic lantern, the instrument accepts multimode (incoherent) light from the telescope focus, which is converted into an array of single-mode outputs.

Since the light from a single-mode fiber, by definition, provides a diffraction-limited source, the single-mode outputs can be lined up into a slit arrangement and fed to a diffraction-limited spectrograph. The spectrograph can thus be in its most compact form (of order 10 cm), regardless of the properties of the input aperture. That is the secret to achieving inexpensive compact instruments on next-generation ELTs, planet rovers and space telescopes.

Microspectrographs have extraordinary potential. A diffraction-limited spectrograph's resolving power (its periodic spacing divided by its line thickness) is given by $R = \lambda/\delta\lambda = mN$ where m is the order of interference and $N$ is the interference grating's finesse—essentially the number of combining beams produced in the spectrograph, or the number of tracks in the PIMMS array waveguide grating. A well-optimized PIMMS can be made very small indeed; a grating with 2,000 lines per millimeter, illuminated over an area diameter of 10 mm, can achieve $R = 20,000$ in first order, in an instrument only 10 cm long.

A single-mode photonic spectrograph inherently offers improved precision over traditional fiber-fed spectrographs. In those instruments, used since the 1970s, the light from the star or galaxy of interest is injected into a large multimode fiber and transported to the spectrograph, where the end of the fiber forms part of a pseudo-slit. But the injected starlight excites a large combination of modes in the fiber, appearing at the output face as an undulating pattern of speckles, varying in time as the atmosphere-induced wavefront error changes. Since this pattern is at the spectrograph's input slit, its spatial structure is convolved with the spectral point-spread function of the spectrograph. This unpredictable modal noise corrupts the shape of a spectral element on the detector, resulting in a loss of spectral precision. The single mode fibers of the photonic spectrograph do not have this problem, as each yields a perfect Gaussian beam thanks to the spatial-filtering effect.

Such instruments have been demonstrated in a University of Sydney CubeSat and on telescopes in Australia, Chile and Hawaii. They work with high efficiency and high reliability—so much so that other functions, like complex filtering and ring resonators, are now being integrated into them. For example, scientists at the Joint Space Science Institute, University of Maryland, USA, have recently demonstrated a complex fiber Bragg grating embedded into a 2-D waveguide.

## Future prospects

Integrated photonics' impact on astronomy is just beginning. Some photonic technologies are only now proving their mettle in astronomical applications. One example lies in the production of accurate optical frequency combs for precision spectroscopic applications—specifically, the detection of extrasolar planets via the radial velocity shift that they impose on their parent star.

As a planet orbits a distant star, its mass tugs its parent star backward and forward by a minuscule amount, producing line-of-sight velocity variations of meters or

> In the future, more and more bulk optical systems will be replaced by on-chip components, including variable delay lines, spectral dispersion and even the photodetectors themselves.

even centimeters per second. Detecting these velocities via the Doppler shift of the star's spectral lines requires an extraordinarily precise spectrograph. Large laser frequency combs can produce a precise set of lines with which to calibrate the spectrograph in real time, but their huge expense and complexity limits their use to only the largest observatories. Cheap, reproducible photonics, however, offer alternative solutions. One is via a fiber etalon, in which the etalon cavity is constructed within a single-mode fiber or waveguide. The small, monolithic design of such an etalon means it can be easily tuned in a closed loop via simple integrated heating elements, and is small and robust enough to be deployed at telescopes worldwide. Another approach is to use a compact ring resonator on a photonic chip to produce such a frequency comb.

Another emerging application is the use of fiber Bragg gratings to filter out problematic atmospheric spectral lines. When observing the spectra of faint, distant galaxies, signals can be badly contaminated by the relatively bright emission lines (such as OH lines) from our own atmosphere. To mitigate this, extremely accurate optical filters can be constructed by laser-writing a pattern of refractive-index modulations into a single-mode fiber core. Fresnel reflections occur at each modulation, and tuning their spacing and depth thus allows a precisely chosen set of narrow wavelengths to be reflected rather than transmitted through the fiber. This, in turn, permits the known set of troublesome OH emission lines, for example, to be filtered out before reaching the spectrograph—as demonstrated in the GNOSIS and PRAXIS instruments at the Anglo-Australian Telescope, Siding Spring Observatory, Australia.

Although not yet in wide use, photonic technologies can accurately analyze the polarization state of light. Polarimetry is an important tool in astronomy, as light becomes polarized by gas and dust around stars and as it travels through the interstellar medium. Waveguides and couplers can be designed to selectively transmit and select specific polarization states, and so photonic-polarimetric instruments are not far off.

Moreover, photonic technologies may allow the measurement of completely new types of astronomical signals, previously inaccessible to astronomers. One example is that of photon orbital angular momentum (OAM), which measures the field spatial distribution of the beam distinct from its polarization state. Photonic devices are now being tested to detect this information, although whether the OAM signal survives astronomical distances is controversial.

Finally, the many individual photonic devices discussed here not only offer important benefits for astronomy, but can be easily combined and reproduced in complex but stable configurations. One could imagine a slew of photonic lanterns, fiber Bragg gratings and spectrographic elements (such as AWGs) built onto a single chip—and the same applies to interferometry. This is the future of astronomical instrumentation once all telescopes achieve the goal of being truly diffraction-limited.

We predict that more and more bulk optical systems will be replaced by on-chip components, including variable delay lines, spectral dispersion and even the photodetectors themselves. Devices such as superconducting, photon-counting, energy-resolving microwave kinetic inductance detectors (MKIDS) may be integrated with the photonic-chip package, with the whole assembly operating cryogenically. This will allow ultra-stable, highly sensitive and easily replicable interferometric instruments to be deployed across the world—enabling the direct imaging of planets around distant stars and the explosive environments of supermassive black holes at the heart of distant galaxies. OPN

Barnaby Norris and Joss Bland-Hawthorn (jbh@physics.usyd.edu.au) are with Sydney Astrophotonic Instrumentation Labs, University of Sydney, Australia.

### References and Resources

- J. Bland-Hawthorn and P. Kern. "Astrophotonics: A new era for astronomical instruments," Opt. Express **17**, 1880 (2009).
- R.M. Roettenbacher et al. "No Sun-like dynamo on the active star ζ Andromedae from starspot asymmetry," Nature **533**, 217 (2016).
- J. Bland-Hawthorn and S.G. Leon-Saval. "Astrophotonics: Molding the flow of light in astronomical instruments," Opt. Express **25**, 15549 (2017).